\documentclass[conference]{IEEEtran}
\IEEEoverridecommandlockouts
\usepackage{cite}
\usepackage{amsmath,amssymb,amsfonts}
\usepackage{algorithmic}
\usepackage{graphicx}
\usepackage{textcomp}
\usepackage{xcolor}
\usepackage{caption}
\usepackage{subcaption}

\usepackage{tikz}
\usetikzlibrary{arrows.meta,positioning}

\usepackage{subcaption} 
\usepackage{multirow}
\usepackage{hyperref}
\usepackage{dirtytalk}
\usepackage[flushleft]{threeparttable}
\def\BibTeX{{\rm B\kern-.05em{\sc i\kern-.025em b}\kern-.08em
    T\kern-.1667em\lower.7ex\hbox{E}\kern-.125emX}}
\usepackage{array}
\newcolumntype{C}[1]{>{\centering\arraybackslash}m{#1}}
\usepackage{float}

 \renewcommand{\baselinestretch}{0.99}
\begin{document}

\title{ Enabling AI-Native Mobility in 6G: A Real-World Dataset for Handover, Beam management, and Timing Advance}
 % or A Real-World Mobility Dataset for AI/ML-Driven Handover and Beam Management in Beyond-5G and 6G Networks
% \author{Mannam Veera Narayana\IEEEauthorrefmark{1} \IEEEmembership{(member, IEEE)}, Rohit Singh \IEEEauthorrefmark{1}, Deepa M.R.\IEEEauthorrefmark{2} AND Radhakrishna Ganti\IEEEauthorrefmark{1}
% \IEEEmembership{(Member, IEEE)}}

\author{\IEEEauthorblockN{Mannam Veera Narayana\IEEEauthorrefmark{1}, Rohit Singh\IEEEauthorrefmark{2},
Deepa M.R.\IEEEauthorrefmark{3}, Radha Krishna Ganti\IEEEauthorrefmark{4}}
\IEEEauthorblockA{Department of Electrical Engineering\\
Indian Institute of Technology
Madras \\ Chennai, India  600036\\
Email: 
\IEEEauthorrefmark{1}narayana@5gtbiitm.in,
\IEEEauthorrefmark{2}rohitsingh@smail.iitm.ac.in,  
\IEEEauthorrefmark{3}deepa\_r@nokia.com,
\IEEEauthorrefmark{4}rganti@ee.iitm.ac.in
}
}

\makeatletter
% \def\ps@IEEEtitlepagestyle{%
% % \def\@oddfoot{\mycopyrightnotice}%
% % \def\@evenfoot{}%
% }
% % \def\mycopyrightnotice{
% % {\footnotesize 979-8-3503-1788-6/24/\$31.00 \copyright2024 IEEE\hfill}
% \gdef\mycopyrightnotice{}
% }

\maketitle
\begin{abstract}
To address the issues of high interruption time and measurement report overhead under user equipment (UE) mobility—especially in high-speed 5G use cases—the use of AI/ML techniques (AI/ML beam management and mobility procedures) have been proposed. These techniques rely heavily on data that are most often simulated for various scenarios and do not accurately reflect real deployment behavior or user traffic patterns. Therefore, there is an utmost need for realistic datasets under various conditions. This work presents a dataset collected from a commercially deployed network across various modes of mobility (pedestrian, bike, car, bus, and train) and at multiple speeds to depict real-time UE mobility. When collecting the dataset, we focused primarily on handover (HO) scenarios, with the aim of reducing the HO interruption time and maintaining continuous throughput during and immediately after HO execution. To support this research, the dataset includes timing advance (TA) measurements at various signaling events—such as RACH trigger, MAC CE, and PDCCH grant—which are typically missing in existing works.

We cover a detailed description of the creation of the dataset; experimental setup, data acquisition, and extraction. We also cover an exploratory analysis of the data, with a primary focus on mobility, beam management, and TA. We discuss multiple use cases in which the proposed dataset can facilitate understanding of the inference of the AI/ML model. One such use case is to train and evaluate various AI/ML models for TA prediction.

\end{abstract}


\maketitle

\section{INTRODUCTION}
\IEEEPARstart{M}obility management is particularly challenging in dense deployments and in higher-frequency bands. In these settings, smaller cell sizes, strong directionality, and rapidly varying propagation conditions increase the frequency of handovers and also make the handover less robust. 
As networks densify, the UE must track a larger set of serving and neighbor cells (in 5G NR, multiple beams per cell), which increases measurement load and creates more short-stay handovers and radio link failures.
In addition, 5G-Advanced (5G-A) and 6G use cases will impose increasingly stringent requirements on radio access networks as they evolve beyond traditional voice, video, and mobile broadband services, pushing for high reliability, low latency, and efficient use of radio resources.
AI/ML-enhanced mobility methods can leverage contextual and historical information (e.g., past measurements, mobility patterns, and location/trajectory context) to improve handover reliability, reduce interruption time, avoid short-stay handovers, and decrease measurement and signaling overhead through measurement prediction, selective reporting, and proactive decision making \cite{Manalastas2022_HOPrediction_syntheticdata_adaboost, Zheng2023_RSRPPPrediction_Realdataset_Encoder}.

AI/ML is becoming an important enabler to meet future performance targets, and it is widely expected that 6G will be the first generation in which AI/ML is native to the design of the cellular system. In 3GPP, the AI/ML framework for the air interface studied in Release-18 \cite{R18_3gpp38843} provides the foundation for common enablers in multiple ML-enabled use cases, including mobility-related functions. Building on this baseline, further work targeting ML-optimized inter-cell mobility continues as the specifications evolve toward Release-20, to enable more robust, resource-efficient, and proactive handover operation.

Developing and validating AI/ML-assisted mobility procedures requires diverse and realistic data that reflect operational networks, heterogeneous mobility patterns, and deployment-specific propagation effects.
Prior work on AI/ML for handover (HO) and beam management (BM) can be broadly grouped into: (i) \textit{indirect} approaches that first predict future link quality (e.g., RSRP/RSRQ/SINR) and then infer the likelihood of HO/BM actions; and (ii) \textit{direct} approaches that learn HO/BM decisions or control policies (e.g., via classification or reinforcement learning). In both categories, model performance and generalisability are tightly coupled to the fidelity, diversity, and availability of the underlying datasets.

The indirect approach includes \cite{Lima2023_HOPrediction}, where an LSTM model is trained to predict future RSRP for serving cells from a mix of ns-3 simulations and limited LTE field measurements, and a subsequent classifier to predict HO events. However, using only the serving-cell RSRP does not capture realistic multi-neighbor conditions that largely determine HO outcomes. 
In \cite{Wu2022_RSRPPrediction_syntheticdata_gradientboost}, a combination of real-time and extended simulation data is considered to train a neural network-based regression model to predict RSRP.  
 
Direct decision/control has been explored via reinforcement learning in \cite{YajnaNarayana2020_HO_rl}, but the study considers only serving and one neighbour cell and relies on synthetic data, limiting realism. Beyond HO triggering, \cite{Manalastas2022_HOPrediction_syntheticdata_adaboost} addresses inter-frequency HO prediction and mitigation using simulated and real-time augmented data (with fewer HO event data), highlighting the practical challenge of obtaining sufficient failure examples. In \cite{Mishra2020_ho_prediction}, the authors proposed a novel algorithm to reduce HO failures in 5G using features such as RSRP, BLER, TA, and beam direction. However, the data and the data-capture procedure are missing, hindering reproducibility and benchmarking. Related to measurement reduction, \cite{11388038_Measurement_pred_sim} investigates ML-based measurement prediction for vehicular connectivity using simulated data. Overall, while these studies demonstrate the promise of ML for HO/BM, many rely on simulators or limited/undisclosed field details, and dataset release is often not the primary contribution.

In~\cite{ghoshal2025handover}, the authors analyse how handover-related configurations vary across cells in an operational 5G network in the United States (US) and how these differences impact HO performance over a long-term measurement campaign, highlighting the need for longer duration mobility datasets.  In addition, \cite{Kousias2024_Largedataset} released a large-scale dataset of 4G, NB-IoT, and 5G non-standalone (NSA) measurements collected over multiple weeks across two mobile network operators. While valuable for studying coverage and end-user performance at scale, it is not specifically tailored to AI/ML mobility procedure design: it does not provide UE-side mobility-event-centric traces for 5G standalone (e.g., labelled HO/re-establishment segments), beam-level serving-cell measurements, or timing-advance (TA) updates aligned with HO dynamics. However, recent work in \cite{Dai_ltm_ta2026} highlights the need for early TA for uplink synchronisation in a pre-configured mobility procedure (e.g., L1/L2-triggered mobility), where a set of candidate cells is configured for the HO.

To address the gaps mentioned above in existing works, we present a real-world measurement dataset on the UE side, collected from a commercially deployed cellular network across multiple mobility modes and speeds. The dataset is designed to support the development of AI/ML models and benchmarking for handover and beam management (BM), as well as broader mobility-related tasks such as mobility-state inference, measurement reduction, and proactive beam/HO control. In addition to releasing the data, we document the end-to-end measurement campaign, experimental setup, and the data acquisition, extraction, and pre-processing pipeline, and we provide an exploratory analysis that illustrates how the dataset can be used for reproducible evaluation toward 5G-Advanced and 6G AI-native mobility.

The key contributions of this work are as follows.
\begin{itemize}
    \item \textbf{UE-side real-world dataset for mobility research:} a feature-rich dataset capturing HO and BM dynamics across multiple mobility modes and speeds, with GPS, timestamps, and TA information to support mobility-event-centric learning and analysis.
    \item \textbf{Reproducible data pipeline:} a detailed description of the measurement campaign and experimental setup, along with the full data extraction, cleaning, and pre-processing workflow.
    \item \textbf{Exploratory analysis for HO/BM/TA:} an exploratory study of handover, beam management, and timing-advance behaviors in the collected traces, providing insight into practical mobility dynamics.
    % \item \textbf{Positioning against prior datasets and methods:} a comparative discussion of related AI/ML-based HO/BM work and existing 4G/5G measurement datasets, clarifying the unique capabilities and limitations addressed by our release.
    \item \textbf{New TA-aware use case:} a concrete use case proposal for timing-advance prediction around handover events to support early synchronization.
\end{itemize}
                            
\section{Experimental Details}
\subsection{Measurement setup}
\begin{figure}[t]
    \centering
    \begin{tikzpicture}[node distance=7mm and 8mm, font=\footnotesize, >=Latex]
        \node[draw, rounded corners, align=center, minimum width=0.26\linewidth, minimum height=8mm] (ue) {UE\\Samsung A53 5G};
        \node[draw, rounded corners, align=center, right=of ue, minimum width=0.22\linewidth, minimum height=8mm] (usb) {USB-C\\data cable};
        \node[draw, rounded corners, align=center, right=of usb, minimum width=0.28\linewidth, minimum height=8mm] (pc) {Laptop with \\QXDM};

        \draw[->] (ue) -- (usb);
        \draw[->] (usb) -- (pc);

        \node[draw, rounded corners, align=center, below=of pc, minimum width=0.28\linewidth, minimum height=7mm] (log) {Captured\\log files};
        \draw[->] (pc) -- (log);

        \node[align=center, below=of usb, text width=0.55\linewidth] {};
    \end{tikzpicture}
    \caption{Block diagram of the experimental setup used for UE-side log capturing.}
    \label{fig:exp_setup_block}
\end{figure}
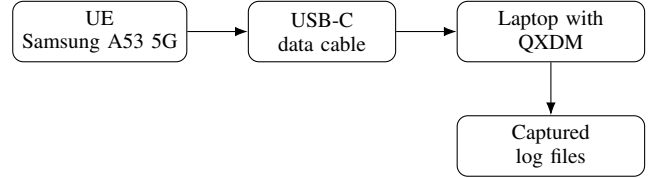

The experimental setup consists of a commercial off-the-shelf user equipment (Samsung A53 5G) connected to a laptop running Qualcomm eXtensible Diagnostic Monitor (QXDM) through a USB-C data cable, enabling real-time collection of UE-side diagnostic logs during RRC-connected operation. As shown in Fig.~\ref{fig:exp_setup_block}, the UE--cable--PC chain forms the complete measurement setup, which is moved to capture data in various mobility modes.

\begin{figure}[t]
    \centering
    
    % First row
    \begin{subfigure}[b]{0.49\linewidth}
        \centering
        \includegraphics[width=\linewidth]{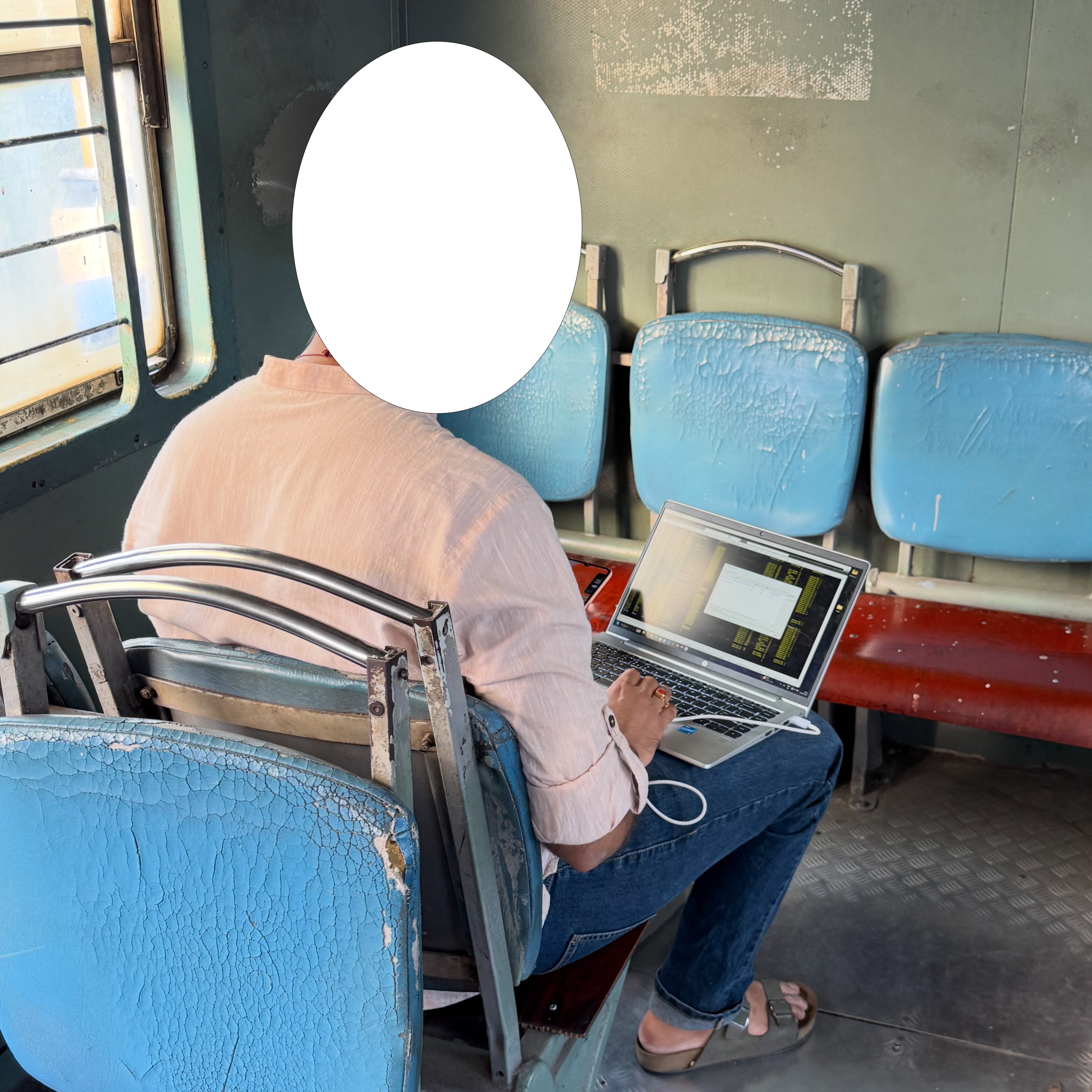}
        \caption{}
        \label{fig:Train}
    \end{subfigure}
    \hfill
    \begin{subfigure}[b]{0.49\linewidth}
        \centering
        \includegraphics[width=\linewidth]{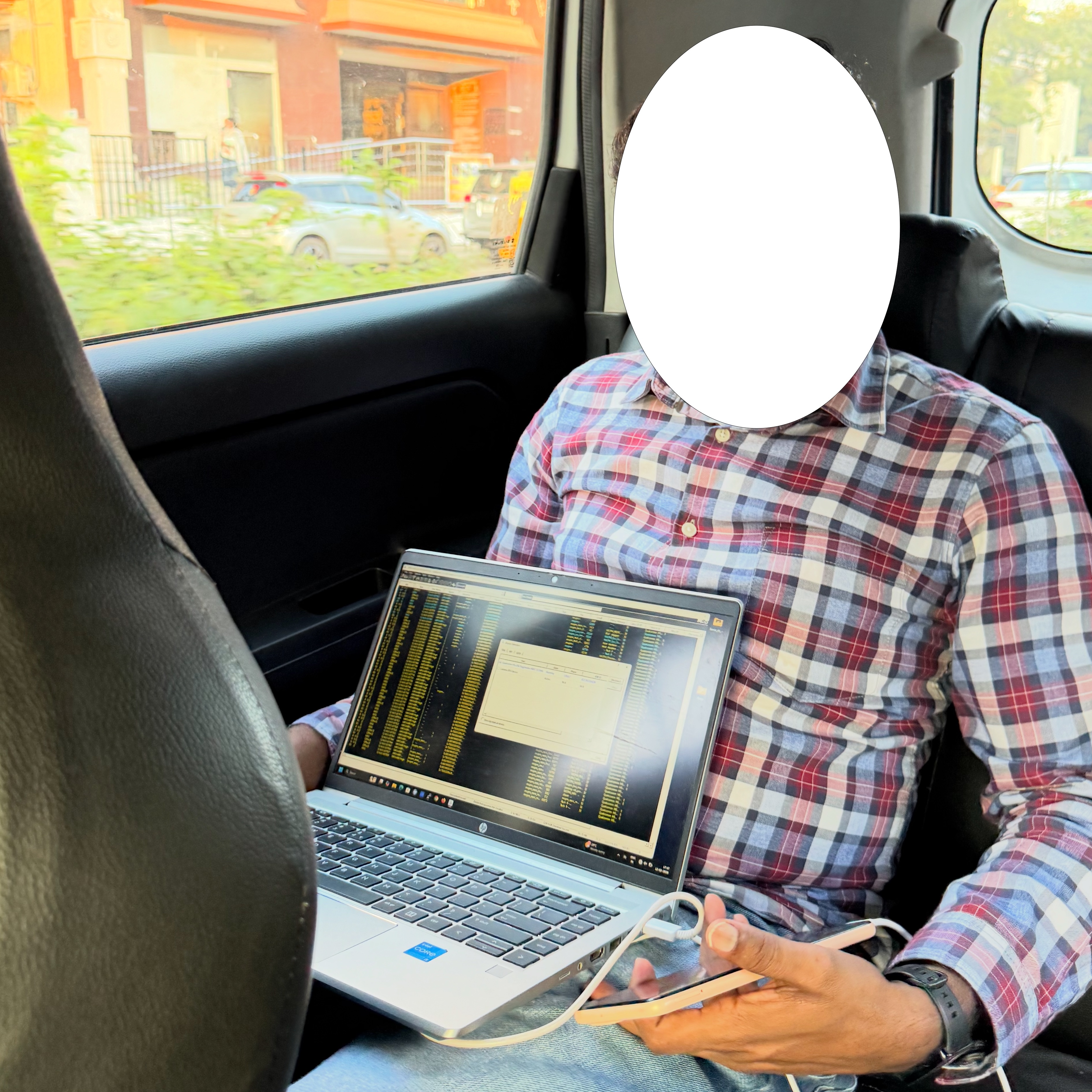}
        \caption{}
        \label{fig:Car}
    \end{subfigure}
    
    \vspace{0.3cm}
    
    % Second row
    
    \begin{subfigure}[b]{0.49\linewidth}
        \centering
        \includegraphics[ width=\linewidth]{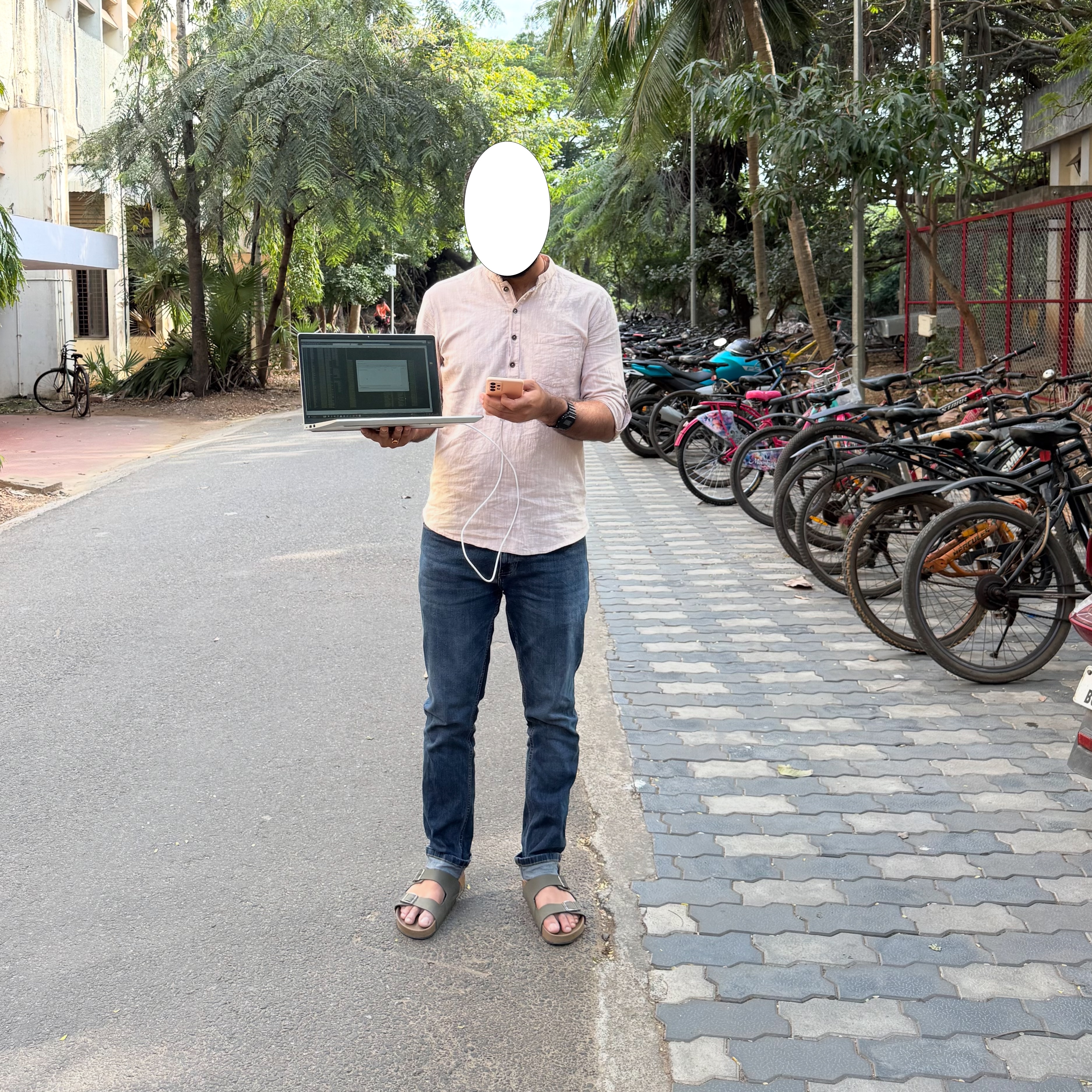}
        \caption{}
        \label{fig:Walk}
    \end{subfigure}
    \hfill
    \begin{subfigure}[b]{0.49\linewidth}
        \centering
        \includegraphics[ width=\linewidth]{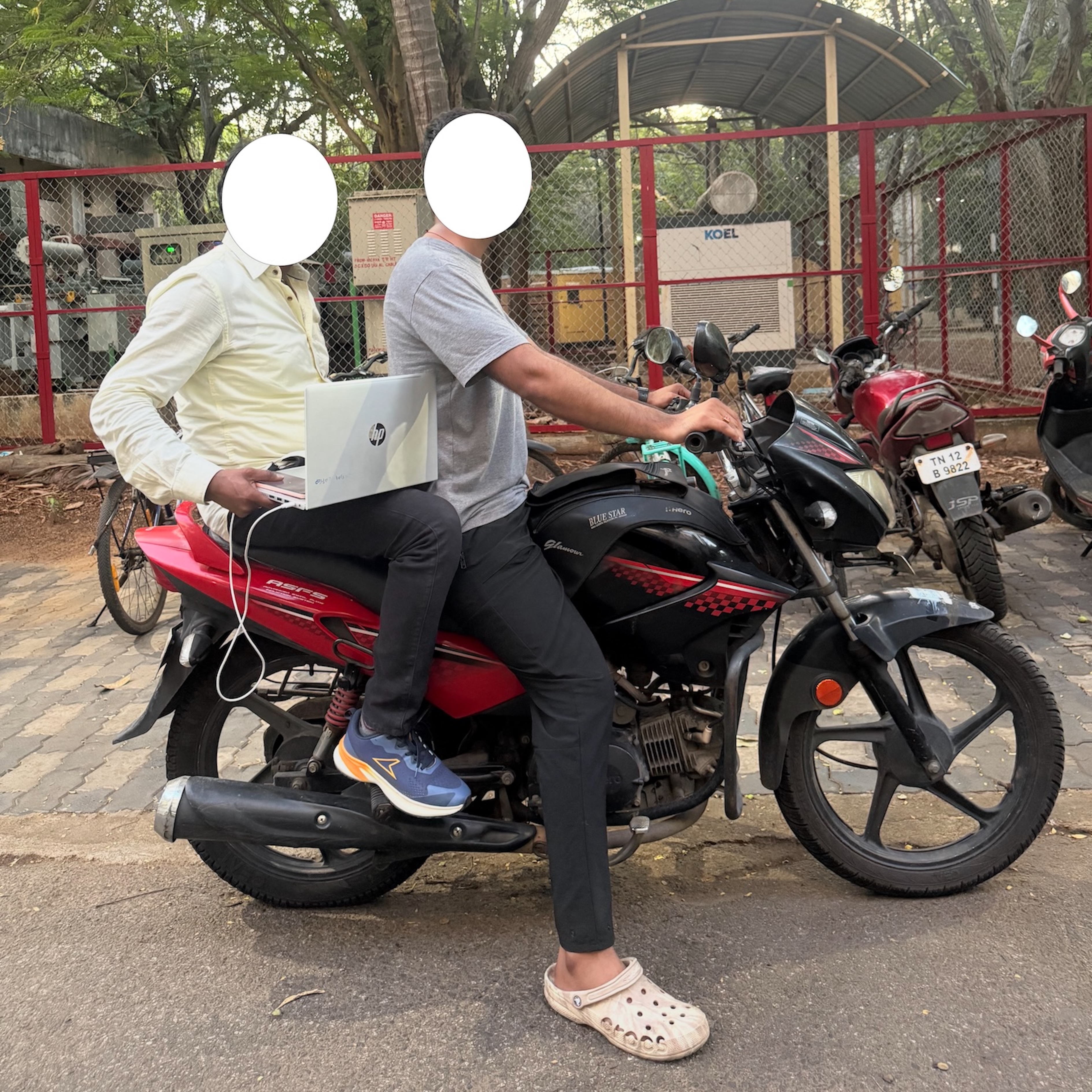}
        \caption{}
        \label{fig:Bike}
    \end{subfigure}
    
    \caption{Experimental setup and data collection in various modes (a) train (b) car (c) walking (d) bike}
    \label{fig: Exp_setup} 
\end{figure}
\subsection{Measurement campaign}
The measurement campaign was conducted in Chennai, India, over several months, from November 2025 to March 2026. Measurements were taken in multiple parts of the city, in and around IIT Madras campus. The campaign followed an opportunistic approach based on daily movement and was carried out in different modes of mobility, walking, car, train, bus and bike, as shown in Fig.~\ref{fig: Exp_setup}. The average speed of the different modes of mobility is as follows. 

        \begin{itemize}
            \item Walking: 4.5 $km/h$
            \item Car: 45 $km/h$
            \item Bike: 30 $km/h$
            \item Bus: 20 $km/h$
            \item Train: 65 $km/h$
        \end{itemize}

\begin{figure}[ht!]
    \centering
    \includegraphics[width=0.43\textwidth]{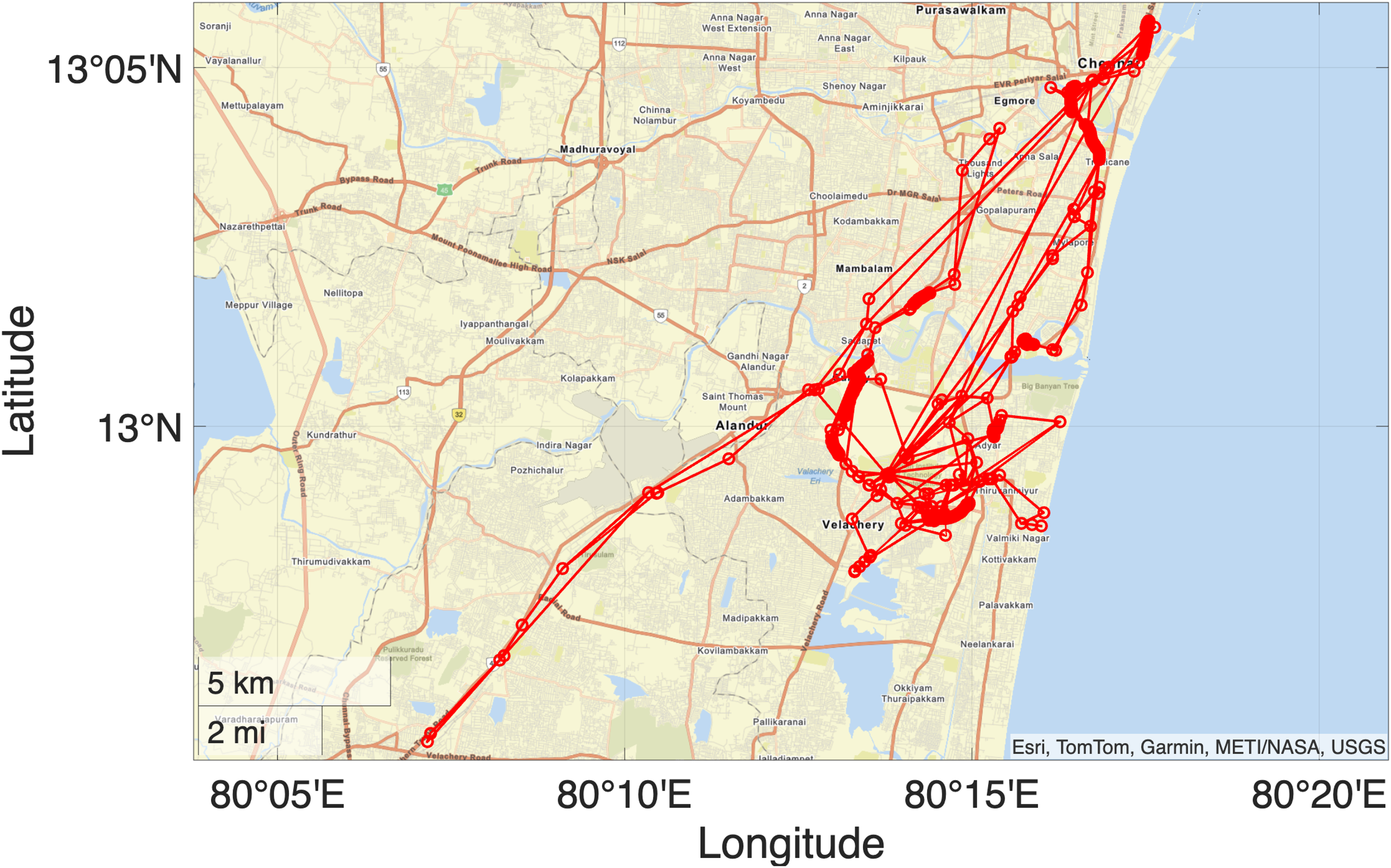}
    \caption{Measurement campaign on Google Maps. The red track indicates the route covered in all the modes of mobility.}
    \label{fig: capture_locs}
\end{figure}

Data collection was not performed every day throughout the campaign. Instead, captures were taken on non-consecutive random days. Moreover, even on days when measurements were taken, the logging was not continuous throughout the day. The campaign used an episodic capture strategy: each capture episode lasted up to $30$~minutes. The upper bound of $30$~minutes is dictated by the dynamic memory limitations of the logging system. QXDM buffers several samples in the running dynamic RAM during capture, and in our setup, the available buffer capacity supports reliable logging for up to approximately half an hour. After each episode, the captured logs are saved in the system's main memory before initiating the next capture. Apart from these memory-driven constraints, no additional technical restrictions motivated the episodic nature of the measurements. The route of the measurement campaign is shown in Fig.~\ref{fig: capture_locs}. The relevant network and UE configurations used during the campaign (including the measurement and reporting periodicity) are summarized in Table~\ref{tab:gnb_config}.

\begin{table}
    \centering
    \caption{ Table showing the configuration used by the commercial network gNB}
    \begin{tabular}{|c|c|}\hline
         Parameter&  Value\\\hline
         Band&  n78\\\hline
         % SSB periodicity&  20 msec\\\hline
         Measurement periodicity&  20 subframes\\\hline
         Reporting periodicity&  5120 msec\\\hline
         Subcarrier spacing&  30 kHz\\\hline
         Measurement signal&  SSB\\\hline
 Downlink Bandwidth&100 MHz\\\hline
         Uplink Bandwidth&  100 MHz\\ \hline
 PRACH Format&Format 0\\\hline
    \end{tabular}
    
    \label{tab:gnb_config}
\end{table}

\subsection{Data Collection and Pre-processing}
Each capture run starts with the UE in flight mode to ensure clean initialization. Data collection begins by disabling flight mode and establishing normal network connectivity, after which QXDM logs the UE activities and protocol messages into its running dynamic memory. As described in the measurement campaign, we then move opportunistically along a selected route for up to $30$~minutes while recording logs. At the end of each capture interval, the recording is stopped, and the buffered logs are saved from dynamic memory to the system's main memory. This procedure is repeated multiple times across different routes within Chennai, India, resulting in multiple independent capture files.

\begin{figure}[b]
    \centering
    \includegraphics[width=0.48\textwidth]{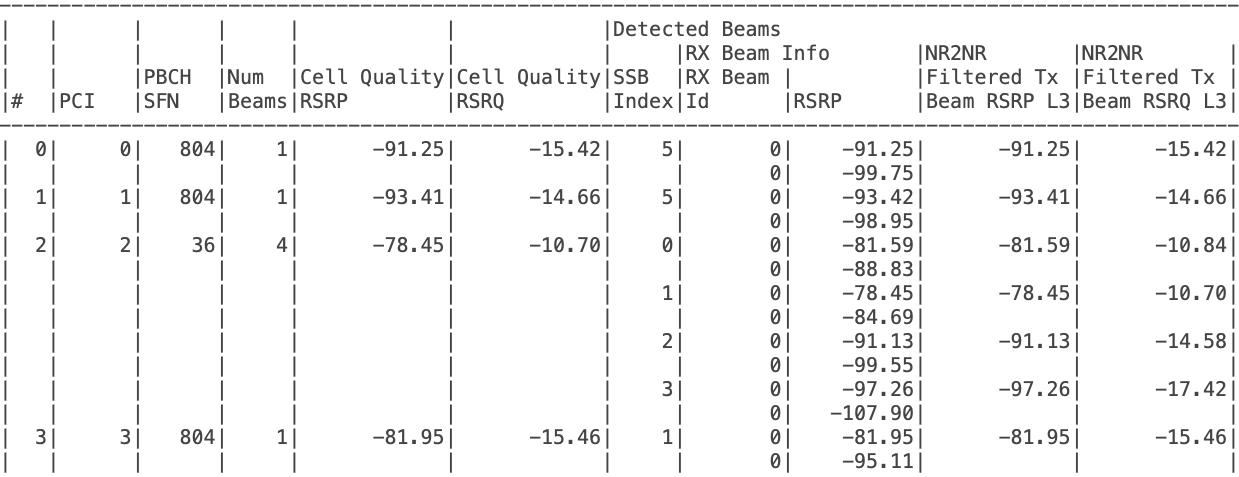}
    \caption{An example measurement report sent by the UE to the gNB and captured by QXDM.}
    \label{fig: meas_rep}
\end{figure}

For pre-processing, the raw QXDM logs are first read into Qualcomm Commercial Analysis Toolkit (QCAT) to filter out only 5G stand-alone (5G SA) messages, which are then exported and stored as text files for offline parsing. A sample measurement report in text file format is shown in Fig.~\ref{fig: meas_rep}. From the 5G SA message subset, we further extract only the packets relevant to radio measurements and RRC signaling to derive the measurement reports, PRACH timing advance (TA) and their corresponding timestamps. Feature extraction is performed using a dedicated Python script, which is provided in the data resources section to support reproducibility. Various features of the dataset are illustrated below.

\begin{itemize}
    \item \textit{Timestamp}: Date and time at which a sample was recorded (format: dd.mm.yyyy and HH:MM:SS.sss).

    \item \textit{PCI}: Physical Cell Identity used by the network to distinguish cells. Each sample includes the PCIs for the serving and neighboring cells. For anonymity, we remap the original PCIs reported on the network to pseudonymous identifiers.

    \item \textit{Location}: Indicates the location of the UE (latitude, longitude). Location timestamps can differ from measurement timestamps by a few milliseconds due to different reporting periodicities; therefore, each measurement is paired with the temporally closest location sample.

    \item \textit{PBCH SFN}: System Frame Number (SFN) on which the SSB measurement was done.

    \item \textit{Number of SSB beams and SSB index}: For the serving cell, the UE reports beam-level measurements across the detected SSB beams; each beam is identified by its SSB index. For neighbor cells, the UE reports only the best beam and the corresponding measurement. Since there is no beamforming on the UE side, the UE uses a single receiving beam.

    \item \textit{RSRP}: Reference Signal Received power at the UE (in dBm)

    \item \textit{RSRQ}: The ratio between reference-signal power and total received power (including interference and noise).

    \item \textit{NR2NR Filtered RSRP}: Moving-average filtered RSRP used for mobility procedures to smooth short-term fluctuations.

    \item \textit{NR2NR Filtered RSRQ}: Moving-average filtered RSRQ used for mobility procedures to smooth short-term fluctuations.

    \item \textit{Timing Advance (TA)}: Uplink timing adjustment commanded by the gNB to align the UE transmissions in time. TA is estimated during random access (RACH) and updated upon re-establishment events such as new connections, handovers, or radio-link failures.
    \item \textit{RACH reason}: The reason why RACH is triggered
    \item \textit{RACH SSB ID}:  The SSB index used for RACH.
\end{itemize}

\section{Exploratory Data Analysis}
In this Section, we present an exploratory analysis of the dataset focusing on handover, beam management, and timing advance.

\begin{figure}[t!]
    \centering
    \begin{subfigure}{0.48\textwidth}
        \centering        \includegraphics[width=\linewidth]{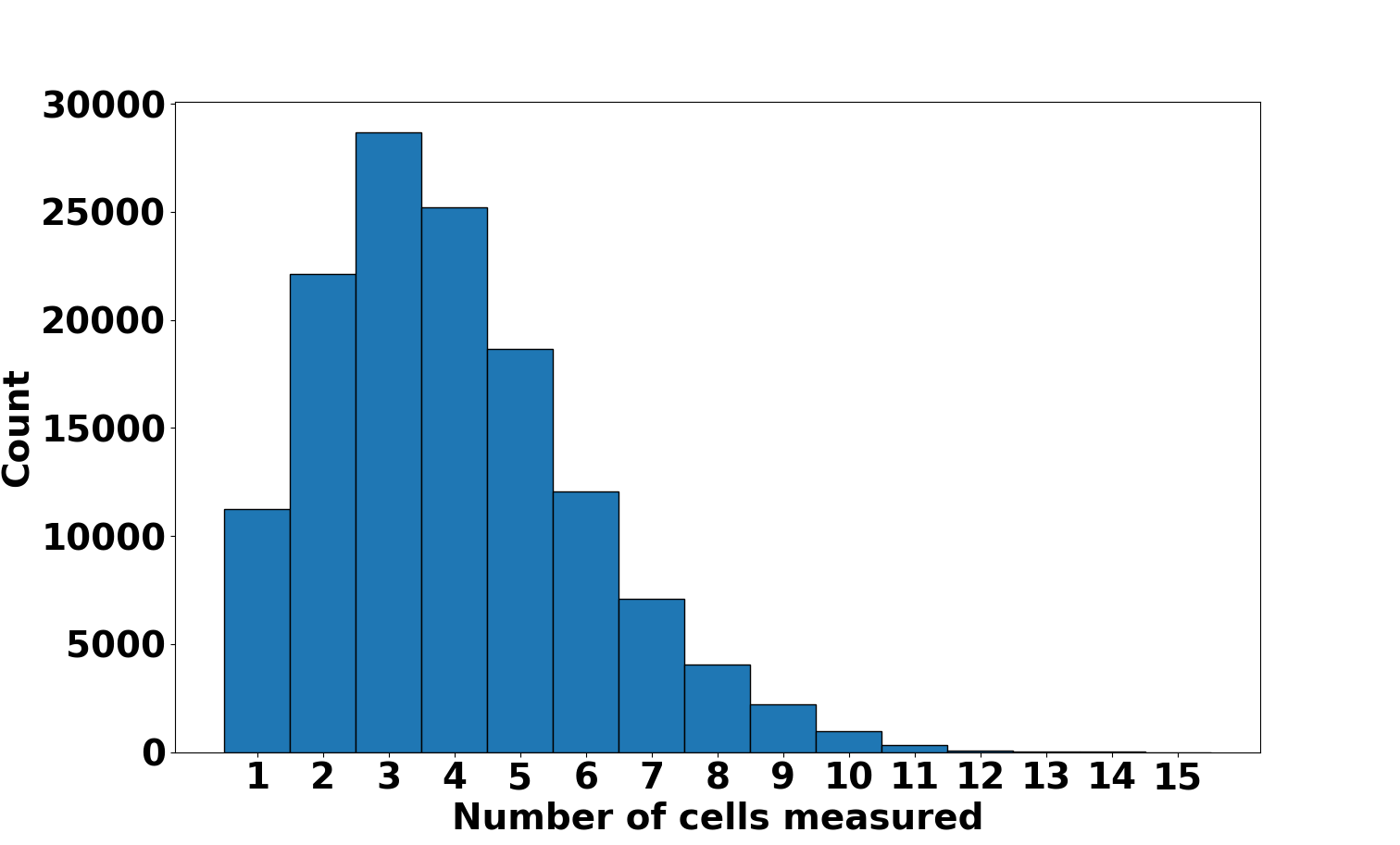}
        \caption{}
        \label{fig: cell_hist}
        \hfill
    \end{subfigure}
    % \hfill
    \begin{subfigure}{0.35\textwidth}
        \centering
        \includegraphics[width=\linewidth]{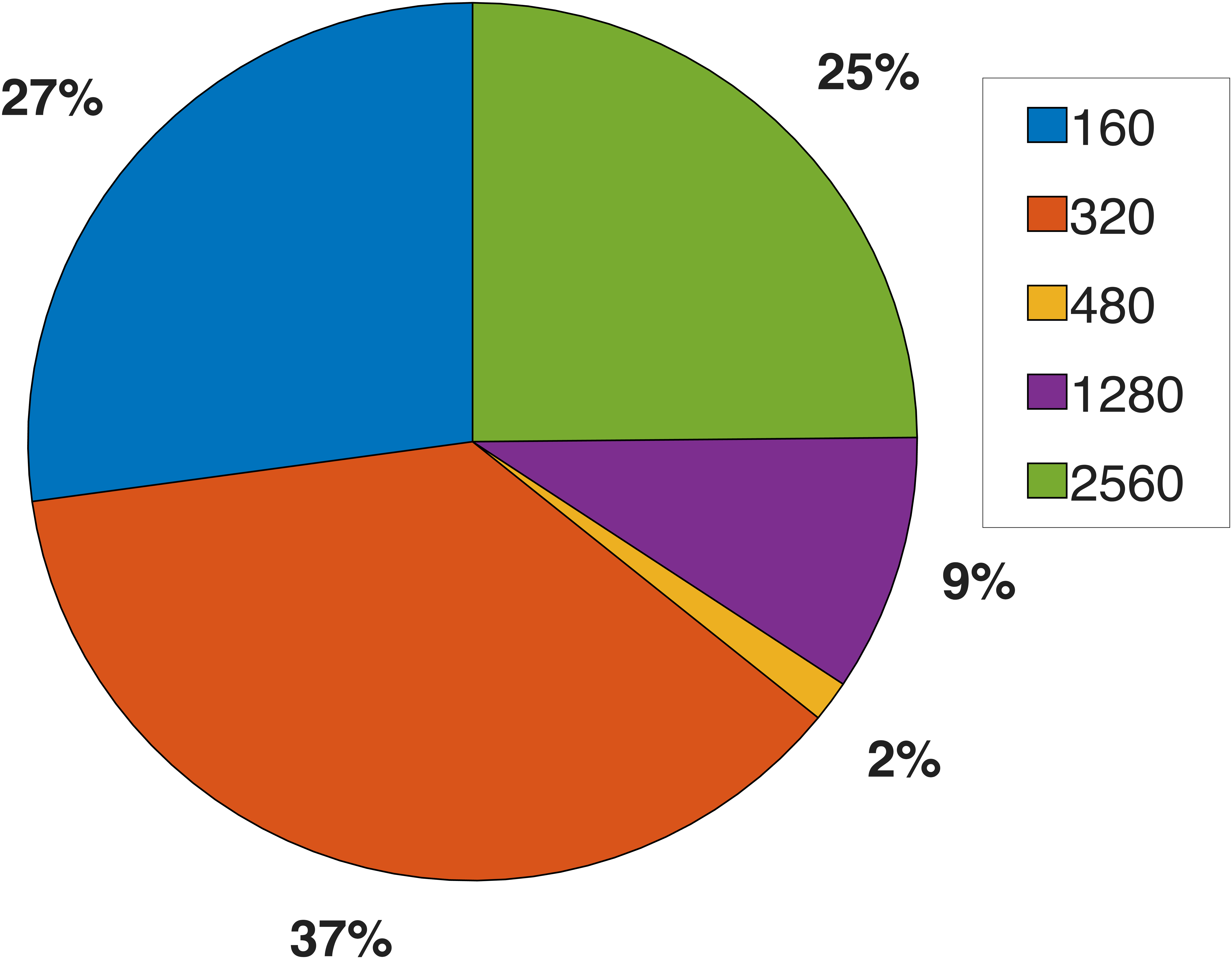}
        \caption{}
        \label{fig:Measurement_time_diff}
    \end{subfigure}
    \caption{UE measurement details during the campaign (a). Distribution of the number of cells measured by the UE.  (b). Distribution of the measurement periodicity of the UE (in \textit{msec}). Note: small clock errors (in the order of microseconds) are grouped into the same bin while computing the distribution.}
\end{figure}

Combining all measurements from episodic captures, the data set contains $1{,}17{,}390$ samples (measurement reports). Each sample comprises multiple measurements, where the total number of measurements depends on (i) the number of cells observed and (ii) the number of beams measured per observed cell. For the serving cell, the UE performs \textit{beam-level} measurements, i.e., it measures one or more Synchronization Signal Block (SSB) beams (identified by the reported beam/SSB index). For neighboring cells, the UE performs \textit{cell-level} measurements, i.e., it captures a single measurement per neighbor cell. Therefore, each sample contains multiple beam measurements for the serving cell and one measurement for each neighbor cell. The total number of measurements per sample is
$M_t = B_s + N_n$,
where $B_s$ is the number of beams measured in the serving-cells and $N_n$ is the number of neighbor cells measured. Fig.~\ref{fig: cell_hist} shows the distribution of the number of cells measured during the campaign; about $96\%$ of the samples UE measured two or more cells. 

The measurement periodicity is not fixed. Distribution of the periodicity of the measurement of the UE shown in fig.~ \ref{fig:Measurement_time_diff}. During the computation of this distribution, we grouped small clock errors (on the order of microseconds) into the same bin.

\begin{figure}[ht!]
    \centering
    \includegraphics[width=0.48\textwidth]{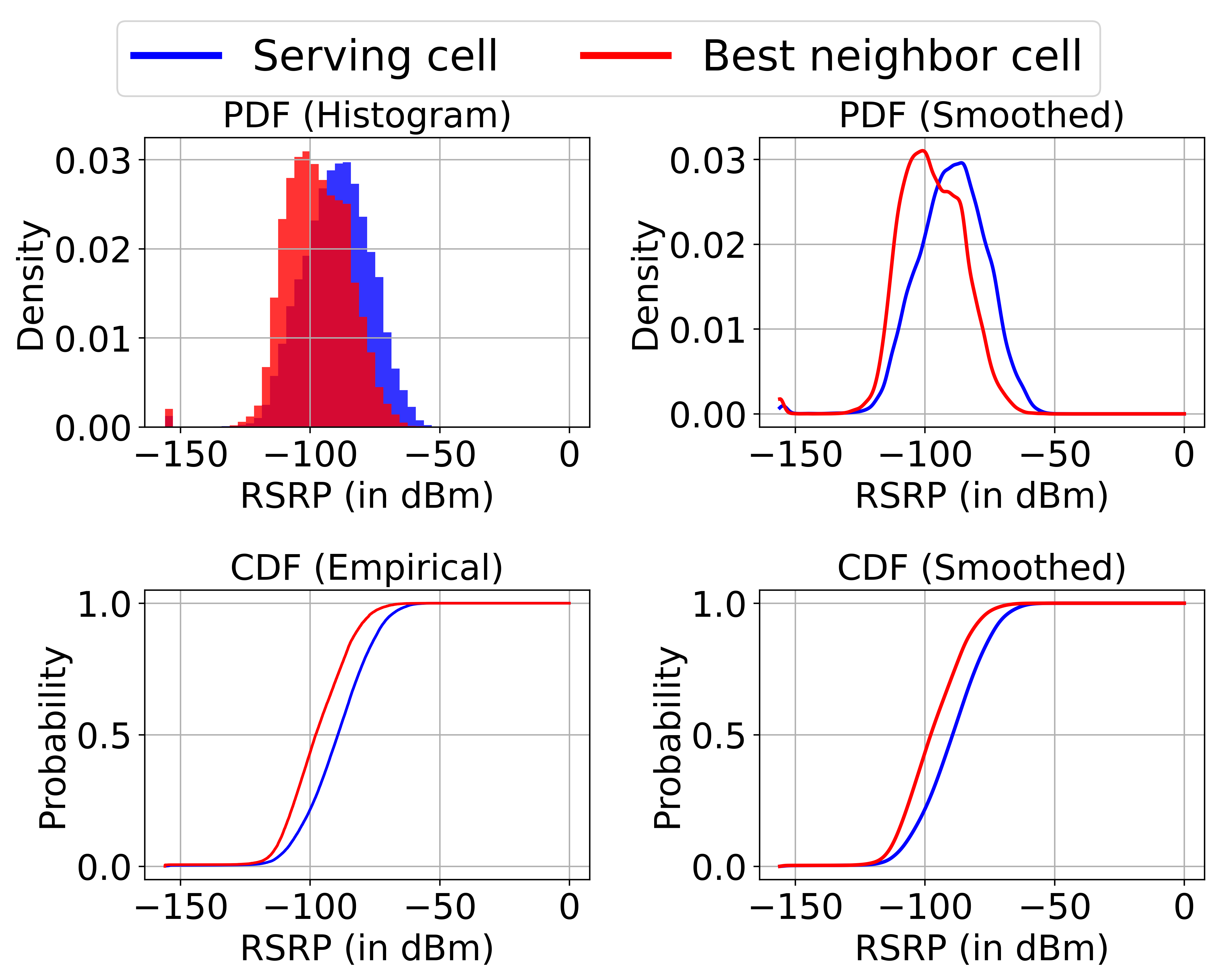}
    \caption{Distributions of the Serving cell RSRP and best neighbor cell RSRP.}
    \label{fig: SC_nbr_RSRP_distr}
\end{figure}

In measurements, RSRP is one of the KPIs that directly affect the quality of service at the UE's end. Fig.~\ref{fig: SC_nbr_RSRP_distr} shows the probability distribution and cumulative distribution of the RSRP of the serving cell and the RSRP of the best neighbour cell. From Fig.~\ref{fig: SC_nbr_RSRP_distr}, we can observe that $80\%$ of the time, the RSRP of the serving cell is above $-100\text{ } dBm$, indicating a good coverage of the 5G signal. The distribution of RSRP for serving cells is very similar to that of T-Mobile and is better than that of AT\&T and Verizon~\cite{ghoshal2025handover} in the US.

\subsection{Handover Analysis}
The handover analysis (HO) is performed on the data set obtained by merging all measurement campaigns. We focus on the serving-cell changes observed in the measurement reports and relate them to the configured 3GPP NR measurement events.

\begin{figure}[htbp!] 
    \centering
    \includegraphics[width=\linewidth]{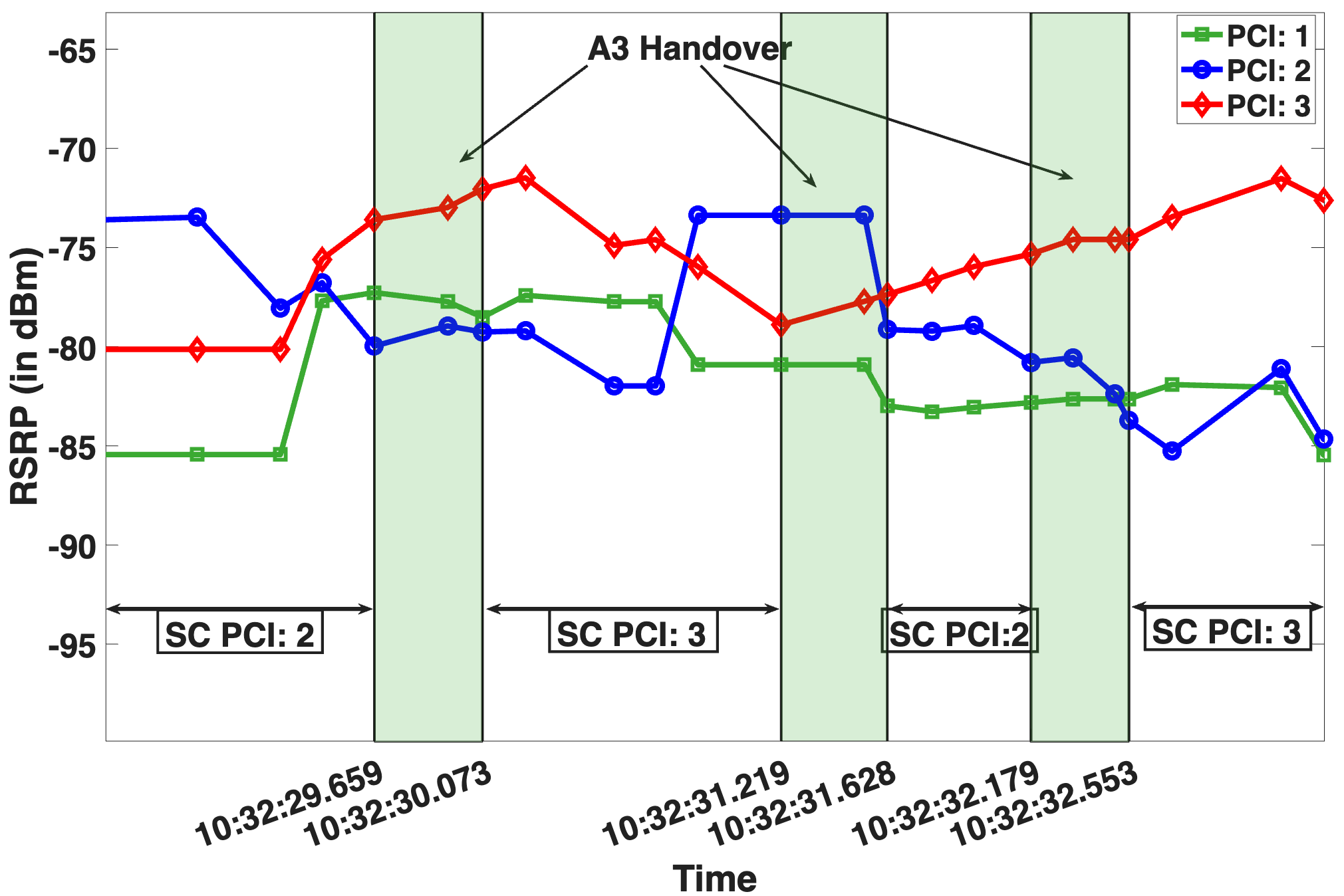}
    \caption{ Sample  A3 handovers in the collected dataset}
    \label{fig:Case1}
    \end{figure}
    
\begin{figure} [htbp!]
        \centering
        \includegraphics[width=\linewidth]{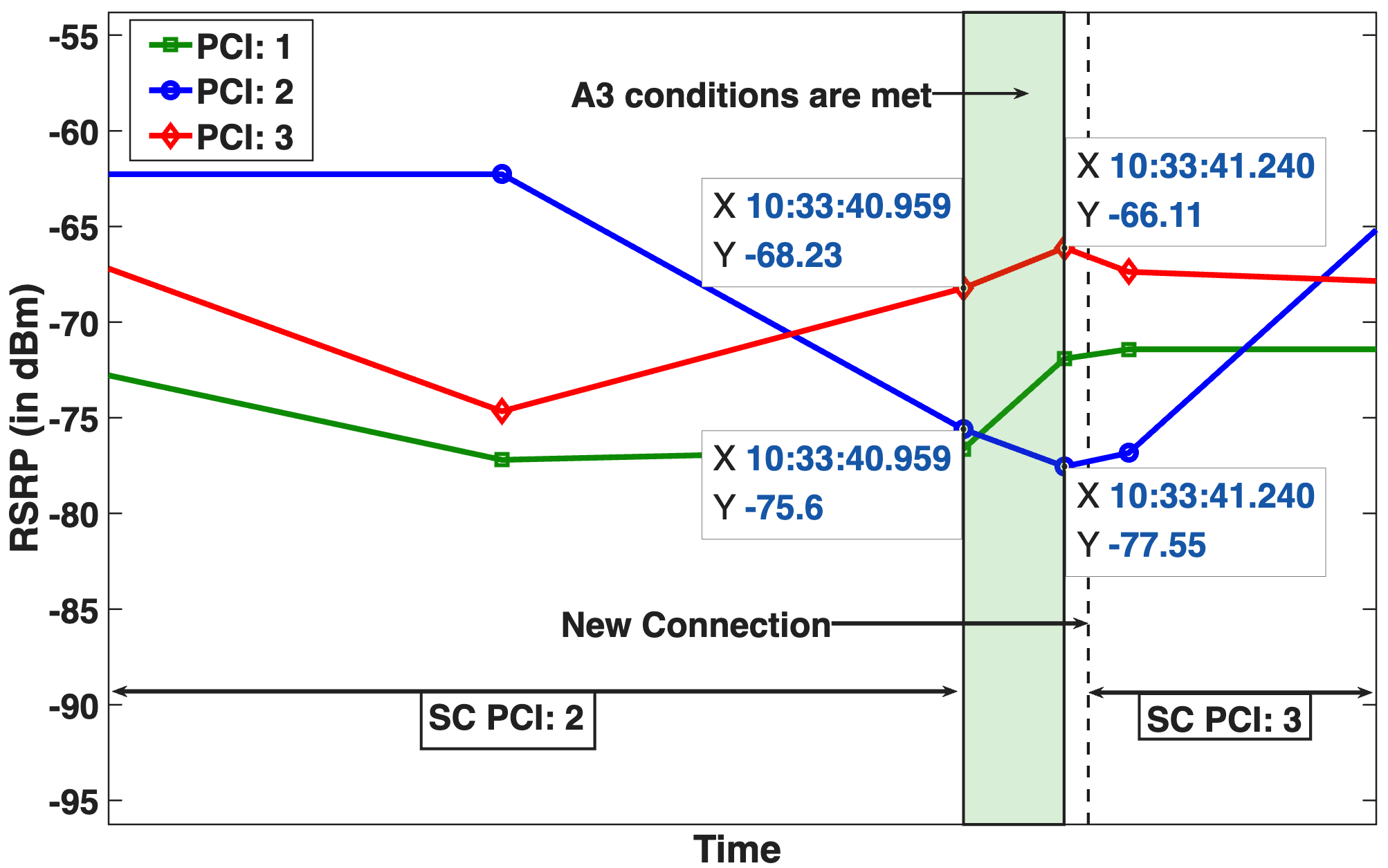}
        \caption{A sample of unsuccessful handover from the collected dataset, despite meeting the HO conditions. Here, SC PCI is the serving cell PCI.}
        \label{fig:Case2}
\end{figure}

As the UE moves, it continuously evaluates the serving cell and neighbor cells and generates measurement reports according to the network configuration. In 5G NR SA, six measurement events are defined (Table~\ref{tab:handover}). In our measurement report, the network is configured with events $A2$ and $A3$; the corresponding parameter settings are summarized in Table~\ref{tab:A2parameters} and Table~\ref{tab:A3parameters}.

\begin{table}[ht!]
    \centering
    \begin{tabular}{|c|>{\centering\arraybackslash}p{0.8\linewidth}|}\hline
         Event & Trigger condition\\\hline
         A1& Serving cell becomes better than absolute threshold\\\hline
         A2& Serving cell becomes worse than absolute threshold\\\hline
         A3& Neighbor cell becomes amount of offset better than PCell/PSCell\\\hline
         A4& Neighbor cell becomes better than absolute threshold\\\hline
         A5& PCell/PSCell becomes worse than absolute threshold1 AND Neighbour/SCell becomes better than another absolute threshold2\\\hline
         A6& Neighbour cell becomes amount of offset better than SCell\\ \hline
    \end{tabular}
    \caption{Measurement events defined in TS 38.331 \cite{3gpp_38_331}}
    \label{tab:handover}
\end{table}

\begin{table}[ht!]
\caption{Event A2 parameters}
    \centering
    \begin{tabular}{|>{\centering\arraybackslash}p{0.2\linewidth}|>{\centering\arraybackslash}p{0.2\linewidth}|>{\centering\arraybackslash}p{0.2\linewidth}|>{\raggedright\arraybackslash}p{0.2\linewidth}|}\hline
         RSRP Threshold&  Hysteresis& Time to Trigger (TTT) &Reporting periodicity\\\hline
         -112 dB&  0 dB& 480 msec&240 msec\\ \hline
    \end{tabular}
    \label{tab:A2parameters}
\end{table}

\begin{table}[ht!]
\caption{Event A3 parameters}
    \centering
    \begin{tabular}{|>{\centering\arraybackslash}p{0.2\linewidth}|>{\centering\arraybackslash}p{0.2\linewidth}|>{\centering\arraybackslash}p{0.2\linewidth}|>{\raggedright\arraybackslash}p{0.2\linewidth}|}\hline
         Offset&  Hysteresis& Time to Trigger (TTT) &Reporting periodicity\\\hline
         3 dB&  2 dB& 256 msec&240 msec\\ \hline
    \end{tabular}
    \label{tab:A3parameters}
\end{table}

From the combined dataset, we observe two serving-cell change scenarios:
\begin{enumerate}
    \item \textbf{Case 1: A3 handover;} The change in the serving cell in the A3 handover occurs whenever the entry condition in \eqref{eq:A3entry} is met and the exit condition in \eqref{eq:A3exit} is not met before the expiration of the Time to Trigger (TTT) timer. In our measurement campaign, it is configured to use RSRP as the primary trigger for HO decisions. Hysteresis and TTT parameters in \eqref{eq:A3entry}, \eqref{eq:A3exit} mitigate spurious triggers (e.g. ping-pong), and the configured network values for these parameters are shown in Table \ref{tab:A3parameters}.  Upon executing a handover, the PCI of the serving cell changes accordingly. Fig.~\ref{fig:Case1} illustrates three HOs based on the A3 event, highlighted in the green strip. Initially, the serving cell is PCI: 2 (the blue line), which is changed to PCI: 3 (the red line) at the first HO, to PCI: 2 (the blue line) at the second HO, and to PCI: 3 at the third HO. In total, $1546$ A3 handovers are recorded in the dataset. From Fig.~\ref{fig:Case1}, we can also observe the ping-pong effect at the second HO, where the serving cell PCI switched from 2 to 3 and then to 2. This can be addressed by adjusting the Hysteresis, TTT, and the $\text{A3}_{offset}$ parameters. However, the optimal values for these parameters depend on several factors such as network load, deployment scenarios, etc. \cite{ghoshal2025handover} 
    \begin{align} 
    \text{RSRP}_{nbr}  > \text{RSRP}_{SC} + \text{A3}_{offset} + \text{Hysteresis} \label{eq:A3entry} \\ 
    \text{RSRP}_{nbr} < \text{RSRP}_{SC} + \text{A3}_{offset} - \text{Hysteresis}  \label{eq:A3exit}
    \end{align}
    Where $\text{RSRP}_{nbr}$ is the RSRP of the neighbor cell, $\text{RSRP}_{SC}$ is the RSRP of the serving cell, $\text{A3}_{offset}$  is the offset configured by the network.
    \item \textbf{Case 2: A3 condition met, but no immediate serving cell change;} In some instances, although the A3 HO conditions are met by one (or more) neighbors, the serving cell remains unchanged. Fig.~\ref{fig:Case2} shows an example in which a neighbour cell (PCI: 3)  meets the A3 HO conditions, yet the UE stays in the original serving cell (PCI: 2). However, the UE connected to PCI : 3 in a fresh connection instead of Handover. Such behavior can occur due to gNB-side implementation and policy decisions (e.g., load balancing, HO robustness constraints, admission control), or because the neighbor becomes ``eligible'' only briefly. 
    
\end{enumerate}

\subsection{Beam management Analysis}
In addition to cell-level mobility (serving cell change), 5G NR employs \emph{beam management} to maintain the best beam within the serving cell. In the dataset, beam-level measurements are also available for the serving cell. For each reporting instant, UE measures the RSRP of the detected SSB beams (identified by the SSB index), and the network updates the serving beam accordingly. Since the beamforming provision is missing on the UE side, the UE uses a single receive beam, while the gNB transmits multiple SSB beams (eight beams in the present deployment).

\begin{figure}[ht!]
    \centering
    \includegraphics[width=0.96\linewidth]{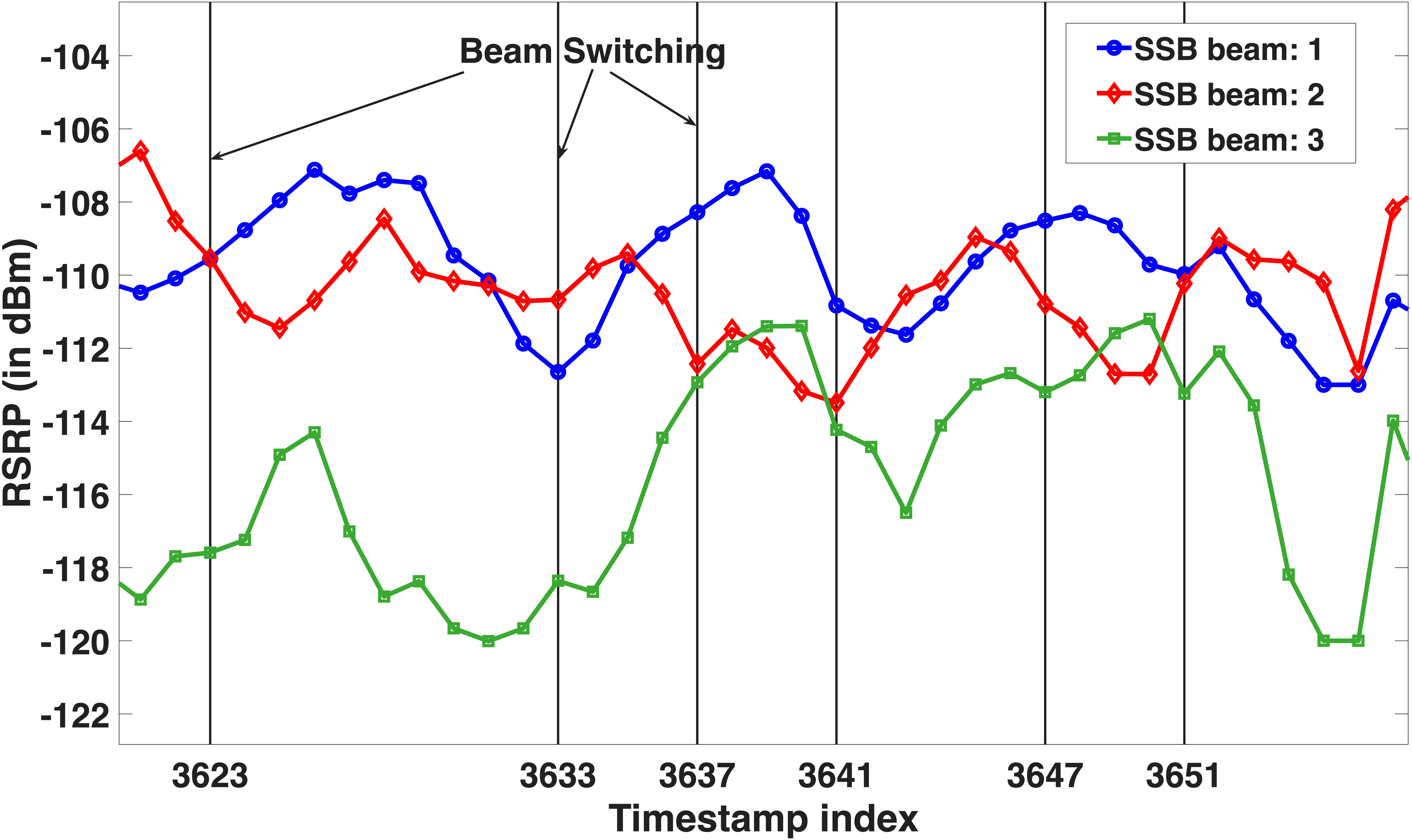}
    \caption{Sample Beam switching from the collected dataset based on RSRP threshold}
    \label{fig:beam_switching}
\end{figure}

Fig.~\ref{fig:beam_switching} illustrates the switching of the beam to which the UE is connected (marked with the vertical lines), whenever the connected beam RSRP falls below a threshold less than the other beams measured by the UE. Across the dataset, the UE measures an average of $3.65$ beams per serving cell. The RSRP threshold for beam switching and the L1 and L3 filtering are vendor-specific. This beam-level data can support future works on beam selection, beam stability, and prediction.

\subsection{TA Analysis}
 TA aligns the UE uplink transmissions at the gNB receiver by compensating for propagation delay. TA is obtained by the network from the UE's random access (RA) procedure and can be updated later via timing-advance commands. In the dataset, TA updates are observed primarily at re-establishment points, such as new connections, serving cell changes, and after radio-link recovery, i.e. when uplink time alignment needs to be refreshed.

\begin{figure}[ht!]
    \centering
    \includegraphics[width=0.4\textwidth]{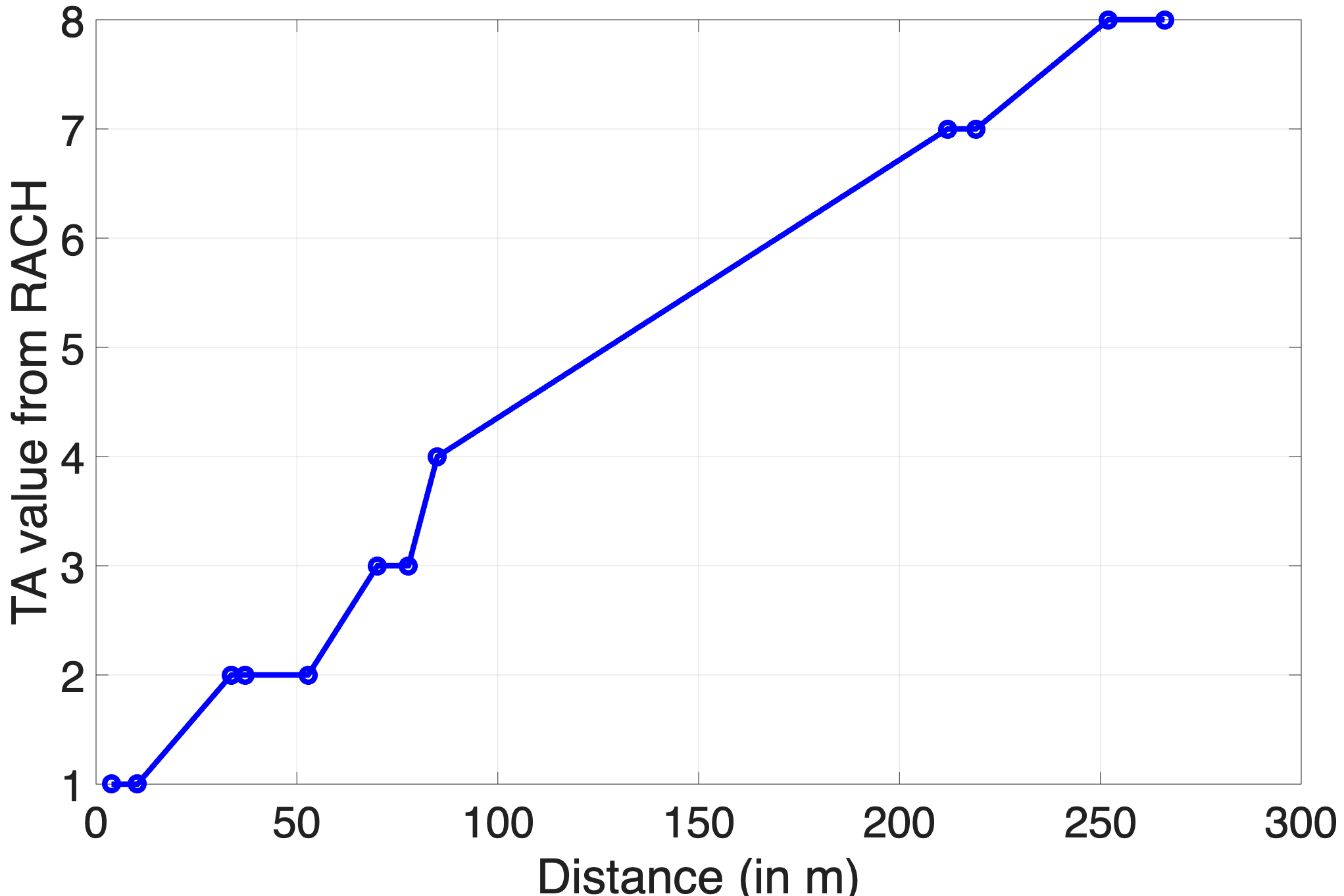}
    \caption{Relation between the gNB-UE distance and RACH TA captured in IIT Madras 5G testbed }
    \label{fig:dist_vs_Ta}
\end{figure}

\begin{figure}[ht!]
    \centering
    \includegraphics[width=0.48\textwidth]{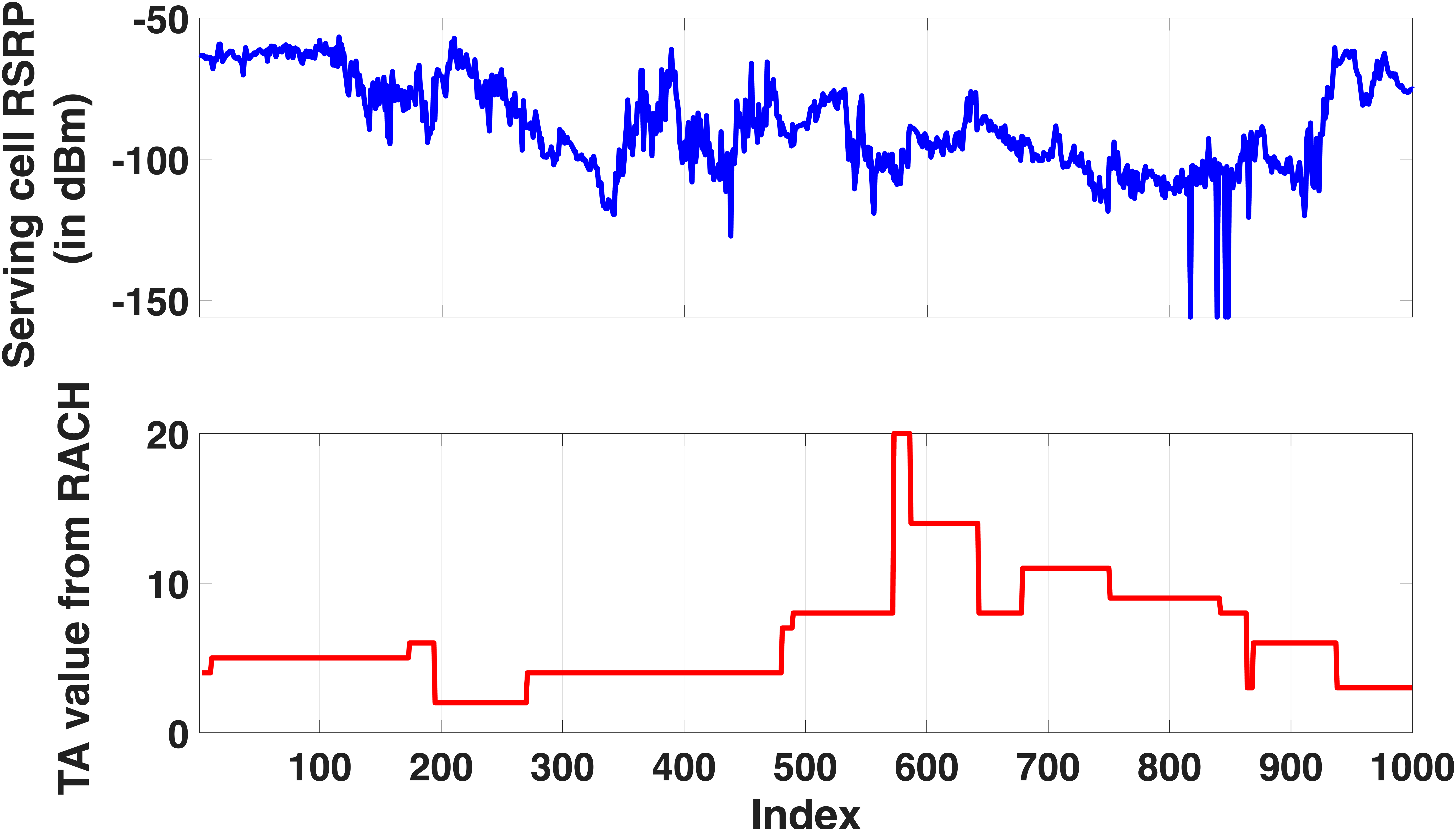}
    \caption{A sample plot for Serving cell RSRP and PRACH TA}
    \label{fig: RSRP_vs_TA}
\end{figure}

Since TA is a quantized (coarse) proxy for round-trip propagation delay, it provides an indirect estimate of the distance between UE and gNB. Although TA does not map uniquely to distance in multipath or non-line-of-sight (NLoS) conditions, we generally expect larger TA values when the UE is farther from the gNB and/or experiences weaker link conditions. To confirm the relationship between the distance of the UE from the gNB and the TA, we also conducted an exclusive experiment by triggering RACH multiple times while collecting measurements. This experiment was conducted on the 5G testbed at IIT Madras \cite{5gtbiitm} and the results are presented in Fig.~\ref{fig:dist_vs_Ta}. A higher RSRP also indicates that the UE is closer to the gNB and has less TA, and vise versa. A similar trend is visible in Fig.~\ref{fig: RSRP_vs_TA}, where high RSRP typically coincides with lower TA. In the dataset, RACH is triggered $3,196$ times, of which $1,546$ are due to handover, $ 1,461$ are due to MAC CE command, $177$ are due to PDCCH grant, and $12$ are due to uplink grant.

The joint availability of serving-cell RSRP/RSRQ and TA in the dataset enables the study of distance-aware mobility behaviour and the detection of events such as cell-edge operation and re-synchronisation episodes around handover or connection re-establishment.

\subsection{TA prediction for Early Synchronisation - use case proposal}
In a legacy handover, the UE corrects its timing before starting data transmission with the target cell. The UE transmits a contention-free preamble (\emph{msg1}) to the target cell, and the target cell estimates the TA and returns it to the UE via a Random Access Response (\emph{msg2}). The general random-access procedure increases the handover interruption time. To reduce the handover interruption time, in \emph{Rel-18} \cite{LTM}, L1/L2 Triggered Mobility (LTM)  was proposed. The measurement-coupled TA values in the dataset can support AI/ML work on TA prediction, enabling early UL synchronisation and, in turn, reducing interruptions during the HO. This could be a potential extension of the present work on dataset creation. 

\section{Data Resources}
\label{sec:data_resources}

Data resources will be available upon acceptance of the article for publication.

\section{Conclusion}
In this paper, we present a real-world measurement dataset collected from a commercially deployed 5G network across various mobility modes, including pedestrian, bicycle, car, and train, as well as different speeds. Our goal is to enable AI-driven mobility research for 5G-Advanced and 6G technologies. The data set supports reproducible exploratory analysis and benchmarking for handover and beam management, capturing the irregularities and heterogeneity that are often missing in purely simulated data. In addition to the data set, we provide, HO, BM, and TA analyses of the collected data.  Notably, the simultaneous availability of Timing Advance (TA) and beam measurements enables TA-aware learning, such as early TA prediction for low-interruption mobility, and opens new opportunities to improve latency-critical handover processes. This could be a potential extension of the current work. In future work, we also plan to expand the measurement campaign to include additional frequency bands, multiple network configurations, and environments to enhance the diversity of the dataset.

% \section{Acknowledgments}
\section*{Acknowledgment}
The authors would like to thank Nokia for supporting the analysis and procedural outline, and the Ministry of Electronics and Information Technology (MeitY), Government of India, for providing the testbed, instruments, and software support through the Next-Generation wireless research and standardization project on 5G and Beyond.

% \section*{REFERENCES}
% references section

\bibliographystyle{IEEEtran}

\bibliography{bibfile}

\end{document}